# Desalination due to Electrical Image Forces

J. B. Sokoloff, Physics Department and Center for Interdisciplinary Research in Complex Systems, Northeastern University, Boston, MA 02115.

## Abstract

It will be shown that for a solution of salt dissolved in water in contact with a metallic wall, the concentration of salt ions (both positive and negative) within a few Angstroms of the wall can be large enough to exceed the salt's solubility limit, as a result of electrical image charge forces. In addition, since the dielectric constant of water increases from 2.1 at the wall to 81 at about a nanometer from a solid wall, there will be an attractive image potential near the plane on which this increase of the dielectric constant occurs. The possible existence of these image potentials suggests that the salt can be removed from the water by making salt water flow between an array of parallel solid plates..

Onsager and Samara[1] have shown that electrical image charge forces acting on ions dissolved in water can explain the observed increase of the surface tension due to the presence of dissolved ions. Electrical image charge forces at the interface of a salt solution and air are repulsive. At the interface between the salt solution and a dielectric of higher dielectric constant than that of the solution or a metal, in contrast, the force is attractive. There have been many studies of the effect of electrical image forces on surface tension[2,3] and on how electrical image forces affect phenomena traditionally studied by the Poisson-Boltzmann equation, which does not include these forces[4-7]. It has also recently been shown that electrical image forces may play a role in the attraction of ions to the electrodes in capacitive desalination[8]. Electrical image forces have also been shown to play an important role in the freezing of a confined room temperature ionic liquid[9.10]. Here, the effect of electrical image forces on the distribution of salt ions in a solution that is in contact with a metallic surface is considered. It is shown that for metals that do not have a thick oxide coating or a good deal of roughness, the salt ion concentration near the metallic surface can be very large. If the salt concentration near the wall exceeds the solubility of the salt, it will precipitate out of solution. This result suggests that having salt water flow between an array of closely spaced parallel metallic plates, as illustrated in Fig. 1 for two walls, could provide a method to desalinate the water. This geometry was proposed earlier[11] for a method of desalination based on a force of friction that acts on the ions due to either the excitation of electron states of the walls or Ohm's law heating that results from the fact that the ions create a polarization charge in the walls, which is dragged along with the ions as the salt water flows between the walls. These two proposed methods of desalination have the advantage over the usual method of desalination by reverse osmosis or capacitive desalination, in that there are no filters or electrodes that need to be periodically de-fouled. If the parallel plates are perpendicular to the ground, the salt can just fall out from between the plates under the force of gravity. The focus of this discussion is on solutions of salts in which the salt ions have the same charge magnitude, such as sodium chloride. In the end, there will be a brief discussion of how the picture will be modified for salt ions with difference charge magnitudes.

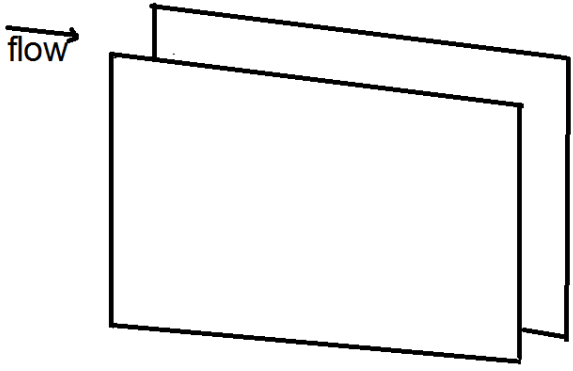

Fig. 1: Illustration of desalination by having salt water flow between two electrically conducting walls.

Consider a salt solution in contact with a metallic wall. Each of the dissolved ions will feel a force $F$, resulting from its electrical image, pulling it towards the wall, of magnitude

$$F = \frac{q^2}{16\pi\varepsilon h^2}, \qquad (1)$$

(at sufficiently low salt concentration to neglect screening), where $h$ is the distance of the ion from the wall, $q$ is the charge of the ion and $\varepsilon$ is the permittivity of water [12]. Since the ions have random positions in the solution, we may assume that the attractive forces between an ion and the image charges due to the other ions cancel out, and hence, can be neglected. In equilibrium, the image forces acting on the ions must be balanced by the force due to osmotic pressure. Consider an infinitesimal region of thickness $dh$ parallel to the metallic wall and having the same length and width as the wall. If the wall has length and width $L$ and $W$, respectively, the equilibrium condition for the ions in this region is

$$LW k_B T \frac{d\rho_i}{dh} dh = -\rho_i LW \frac{dV}{dh} dh, \qquad (2)$$

where $V(h)$ is the potential due to the image charge screened by the conduction electrons and $\rho_i$ is the number density of the ions (i.e., the total number of positive and negative ions) per unit volume. For Debye-Huckel screening $V(h)$ is given by

$$V(h) = \frac{q^2}{16\pi\varepsilon h} e^{-2k_s(h-a)} \qquad (3)$$

where $k_s$ is the inverse Debye-Huckel screening length and $a$ is the distance of closest approach of an ion to the wall. Alternatively, for ions close to the wall, we can use the approach for screening due to Nordholm[13], in which

$$V(h) = \frac{q^2}{16\pi\varepsilon h} \theta(h - \ell_s) \qquad (4)$$

where $\theta(h-\ell_s) = 1$ if $h-\ell_s < 0$ and 0 if $h-\ell_s > 0$. In Ref. 13, $\ell_s = \rho_i^{-1/3}$ is the "correlation hole" radius. Onsager and Samaras[1] also handle screening in this way, but set $\ell_s$ equal to the Debye-Huckel screening length. The left hand side of Eq. (2) is the difference between the force due to the osmotic pressure acting on the side of the region of thickness $dh$ closest and farther away from the wall and the right hand side is the force on the ions in this region due to electrical image forces. This gives us a differential equation for $\rho_i$ which when integrated from $\infty$ to $h$ gives

$$\rho_i(h) = \rho_{i0} \exp\left[\frac{\ell_B}{4h}\right] \qquad (5)$$

if $h < \ell_s$ and $\rho_i \equiv \rho_{i0}$ if $h > \ell_s$, where $\rho_{i0} = \rho_i(h = \infty)$ and $\ell_B = (4\pi\varepsilon_1 k_B T)^{-1} q^2$, the Bjerrum length (where $\varepsilon_1$ is the permittivity near the wall), based on Eq. (4), and

$$\rho_i(h) = \rho_{i0} \exp\left[\frac{\ell_B}{4h} e^{-2k_s(h-a)}\right], \qquad (6)$$

based on Eq. (3), since for both models $V(\infty) = 0$. The value of $a \approx 3 \times 10^{-10} m$. For ion concentrations of the order of that of sea water or smaller, $\ell_s \geq a$. Although in bulk water, $\ell_B = 7 \times 10^{-10} m$, it is known that within about $1 nm$ of a solid surface, $\varepsilon/\varepsilon_0 = 2.1$ [14], compared to 81 in bulk water. This gives a value of $\ell_B = 28 nm$. Thus, we see that for $h$ comparable to $a$, $\rho_i \gg \rho_{i0}$. This implies that the equilibrium ion concentration near the wall can be much larger than its value in bulk water. For a metal surface coated with a $5 nm$ thick oxide coating (as is the case for aluminum, for example), and hence $a \approx 5 nm$, $\rho_i$ is not significantly peaked because $\ell_B/(4a)$ is only equal to 1.4. Doped graphene might be a good candidate because it does not have an oxide coating, and the friction experienced by water as it flows past a graphene surface is very low[15-17].

When the ion concentration exceeds the solubility limit of the salt, the salt will precipitate out. Using $\ell_s = \rho_i^{-1/3}$ for the screening length, this occurs for $h \leq 4.35 nm$, and using the Debye-Huckel screening length, it occurs for values of $h$ of the order of Angstroms. If the plates are perpendicular to the ground, the salt that precipitates out can drop from between the plates, leaving pure water. There would need to be water flowing below the plates (which could be salt water) to wash the precipitated salt away. Also, work must be done against osmotic pressure to maintain the flow of water between the plates by a pump that pulls desalinated water out from between the plates, in order to prevent the desalinated water produced from being pulled back into the salt water behind the plates.

There are some factors not included above that modify this picture. First, the fact that the electronic screening length inside the metal is nonzero has the effect of replacing $a$ by the Thomas-Fermi screening length inside the metal $\ell_{TF}$, if $a$ is much less than $\ell_{TF}$, as discussed in Ref. 10. For a good metal, however, $\ell_{TF}$ is comparable to $a$. Second, if the dielectric constant inside the metal is smaller than that right outside the metal, there will be a repulsive image force that opposes the attractive image described by Eqs. (1-6), due to the change in dielectric constant on entering the metal, since the dielectric constant $\varepsilon/\varepsilon_0$ inside the metal is likely smaller than that of the water just outside the metal's surface[1,10]. If it has a nonzero value smaller than 2.1, the image charge contribution from dielectric polarization, $-q(\varepsilon_2 - \varepsilon_1)/(\varepsilon_2 + \varepsilon_1)$, where $\varepsilon_1$ is the permittivity in the water just outside of the metal and $\varepsilon_2$ is the permittivity inside the metal[12], will have a value between 0 and $q$. Therefore it is likely that it will not completely cancel the image charge $-q$ due to the wall discussed earlier, although it can reduce it. For example, Kornyshev and Vorotyntsev have calculated the net image potential, including both the metallic and dielectric polarization image charges[10] for $\varepsilon_1/\varepsilon_2 = 1.5$. They find that the net image potential is attractive, with a minimum of depth equal to about $0.2q^2/(8\pi\varepsilon_1\ell_{TF})$ located at $h = \ell_{TF}/2$.

Third, there is another image charge force, due to the increase in the dielectric constant as one moves away from the wall. As we shall see, this can also provide an additional desalination mechanism. The dielectric constant of water near a wall is believed to be equal to 2.1 in a layer of thickness $h_0 \approx 1nm$ at the wall, and then to revert to a value comparable to the dielectric constant of bulk water of 81 at $h = h_0$. This will result in an electric image charge force that pulls the ion away from the wall and towards the plane at $h = h_0$ with a force of magnitude $\ell_B' k_B T/(h-h_0)^2$, leading to a potential equal to $-\ell_B' k_B T/[4(h_0-h)]$, where

$$\ell_B' = \frac{q^2}{(4\pi\varepsilon_1 k_B T)} \frac{\varepsilon_3 - \varepsilon_1}{\varepsilon_3 + \varepsilon_1} \approx \ell_B, \qquad (7)$$

where $\varepsilon_3$ is the permittivity for $h > h_0$, for $\varepsilon_3 \gg \varepsilon_1$, and a potential $\ell_B''/[4(h-h_0)]$, where

$$\ell_B'' = \frac{q^2}{4\pi\varepsilon_3 k_B T} \frac{\varepsilon_3 - \varepsilon_1}{\varepsilon_3 + \varepsilon_1} \approx \frac{q^2}{4\pi\varepsilon_3 k_B T} \qquad (8)$$

for $h > h_0$. Although the analysis of the data in Ref. 14 supports a picture in which the permittivity switches over rapidly from $\varepsilon_1$ to $\varepsilon_3$, it is likely that this actually occurs within a distance $\Delta$ of the order of the size of a water molecule. We can account for the nonzero thickness of the region in which the dielectric constant switches from 2.1 near the wall to its bulk value of 81 roughly by replacing $(h-h_0)^{-1}$ by the interpolation formula $\Delta[\Delta^2 + (h-h_0)^2]^{-1}$ with $\Delta$ comparable to the size of a water molecule, which result in the above two terms in the exponential being replaced by

$$-\frac{\ell_B' k_B T (h_0 - h)}{(h-h_0)^2 + \Delta^2} \qquad (9)$$

for $h < h_0$ and

$$\frac{\ell_B"k_BT(h-h_0)}{(h-h_0)^2+\Delta^2} \tag{10}$$

for $h > h_0$. The first of these expressions has a minimum value of $\ell_B'k_BT/(8\Delta)$ at $h = h_0 - \Delta$ and the second has a maximum of $\ell_B"k_BT/(8\Delta)$ (which is $< k_BT$) at $h = h_0 + \Delta$. Thus, the image charge potential due to the change in the permittivity of water on moving away from the wall [Eq. (9)] can be even more attractive than that due to the image charge in the wall. The potential given by Eq. (9) will lead to a second positive term in the argument of the exponential in Eqs. (5,6), resulting in a second peak in $\rho_i$ at $h = h_0 - \Delta$, which if greater than the solubility of salt will lead to salt precipitating out of the water. Furthermore, since there is no physical wall located at $h = h_0$, there is no possibility that salt that precipitates out of solution will be prevented from dropping out from between the plates by adhesive forces exerted by a physical wall. In addition, the existence of an image force resulting from the rapid change in the permittivity of water near a wall is believed to occur for both metallic and nonmetallic walls.

Let us consider how rapidly equilibrium is established, and hence, how rapidly the large values of $\rho_i$ near the metal surface predicted above are established. When the system is not in equilibrium, the difference between the force due to the osmotic pressure on the region of thickness $dh$ considered above and the image charge forces must be equal to the force of friction acting on the ions due to the water as they move towards the wall, before equilibrium is established,

$$LWk_BT\frac{d\rho_i}{dh}dh + \frac{q^2}{16\pi\varepsilon h^2}\rho_i dh LW \theta(h-\ell_s) = \gamma v_i \rho_i LW dh. \tag{11}$$

On the basis of the screening approach of Ref. 13, where $\gamma$ is the friction coefficient for the ions and $v_i$ is their velocity towards the wall. The largest value of $v_i$ occurs initially, when $d\rho_i/dh = 0$ and from the above equation, it is given by

$$v_i = \frac{q^2}{16\pi\varepsilon h^2 \gamma}, \tag{12}$$

for $h$ within $\ell_s$ of the wall. For $h - a = \ell_s = 3A^o$, we obtain $v_i = 8.0 \times 10^{-2} m/s$, and hence, the time to reach the wall $(h-a)/v_i = 0.375 \times 10^{-8} s$. Ions for which $h > \ell_s$ are expected to diffuse into the region with $h < \ell_s$ within a time of the order of $(h-\ell_s)^2/D$, where $D$ is the mean diffusion constant of the ions. If the plates were $1\mu m$ apart, the time that it takes for the ions to diffuse this distance is obtained by dividing the square of this distance $(1\mu m)$ by $D$, which is about $10^{-9} m^2/s$ [18], giving $10^{-3} s$. Since salt crystals that precipitate are expected to be larger than the ions, their rate of diffusion away from the wall or the plane on which the dielectric constant changes is expected to be much slower than that of the dissolved ions in the opposite direction, making them unlikely to re-dissolve.

Although the increase of the dielectric constant of water from 2.1 to 81 as one moves away from the metal plate will oppose the diffusion of ions towards a wall when $h > h_0$ to replace ions that have precipitated out of solution near the wall, but for $h - h_0 \gg \ell_B$, the repulsive image potential for $h > h_0$ resulting from this effect is negligible, and even for $h$ comparable to $h_0$ it is less than $k_B T$, and hence, will have a small effect. Therefore, since the diffusion rate for such large values of $h$ is comparable to what it would be without the repulsive image force near $h = h_0$.

In order to get an idea of the effects of surface roughness, let us model the roughness by hemispheric asperities of radius $r_0$. The force between an ionic charge $q$ and its image in this hemisphere is given by

$$F = \frac{q^2}{4\pi\varepsilon h^2} \frac{1}{(1 - r_0^2/h^2)^2} \left(\frac{r_0}{h}\right) . \qquad (13)$$

Thus, for $h \gg r_0$, the image force is reduced by a factor $r_0/h$. For $h$ comparable to $r_0$, however, the magnitude of the image charge force is not reduced by roughness.

It was shown that for metals that do not have a thick oxide coating or a good deal of roughness, the salt ion concentration near the metallic surface can be very large. If the salt concentration near the wall or near the plane on which the dielectric constant of the water changes from 2.1 near the wall to 81 further away from the wall exceeds the solubility of the salt, it will precipitate out of solution. This result suggests that having salt water flow between an array of closely spaced parallel plates, as illustrated in Fig. 1 for two walls, could provide an effective method to desalinate the water. The present treatment was for salts consisting of a single positive and negative ion of charges with equal magnitudes, such as sodium chloride. For a salt in which the magnitude of the charge on one of the ions is larger than that on the other salt ion, equations (5) and (6) indicate that $\rho_i$ will be larger near the wall or the boundary between the regions of large and small dielectric constant because $\ell_B$ is larger for the higher charge ion, despite the fact that there are more of the lower charge ions in the solution.